\documentclass[aps,prd,twocolumn,superscriptaddress]{revtex4}
\usepackage{epsfig,epsf}
\usepackage{amsmath}
\usepackage{amsthm}
\usepackage{amsfonts}
\usepackage{amssymb}
\usepackage{dsfont}
\usepackage{multirow}
\usepackage{appendix}
\usepackage{slashed}
\usepackage[active]{srcltx}
\usepackage{psfrag}

\begin{document}

\title{Newly observed exotic doubly charmed meson $T^{+}_{cc}$}
\date{\today}
\author{S.~S.~Agaev}
\affiliation{Institute for Physical Problems, Baku State University, Az--1148 Baku,
Azerbaijan}
\author{K.~Azizi}
\affiliation{Department of Physics, University of Tehran, North Karegar Avenue, Tehran
14395-547, Iran}
\affiliation{Department of Physics, Do\v{g}u\c{s} University, Dudullu-\"{U}mraniye, 34775
Istanbul, Turkey}
\author{H.~Sundu}
\affiliation{Department of Physics, Kocaeli University, 41380 Izmit, Turkey}

\begin{abstract}
In this work, we treat the newly observed doubly charmed four-quark state $%
T_{cc}^{+}$ as an axial-vector tetraquark with content $cc\overline{u}%
\overline{d}$, and calculate its spectroscopic parameters and width. The
mass and current coupling of the tetraquark $T_{cc}^{+}$ are found by means
of the QCD two-point sum rule method by taking into account quark, gluon and
mixed condensates up to dimension $10$. The width of the $T_{cc}^{+}$ is
evaluated using partial widths of decay processes $T_{cc}^{+} \to \widetilde{%
T}\pi ^{0}$ and $T_{cc}^{+} \to T_{cc;\overline{u}\overline{u}}^{0}\pi ^{+}$%
, where $\widetilde{T}=cc\overline{u}\overline{d}$ and $T_{cc;\overline{u}%
\overline{u}}^{0}$ are scalar tetraquarks. To compute the partial width of
the first process, we apply the QCD three-point sum rule approach and
extract numerical value of the strong coupling $g$ that corresponds to the
vertex $T_{cc}^{+}\widetilde{T}\pi ^{0}$. The width of the second decay is
estimated using isospin symmetry and the prediction obtained for the first
channel. Our results for the mass $m=(3868\pm 124)~\mathrm{MeV}$ and width $%
\Gamma=(489\pm 92)~\mathrm{keV}$ of the tetraquark $T_{cc}^{+}$ are in a
nice agreement with recent measurements of the LHCb collaboration.
\end{abstract}

\maketitle


\section{Introduction}

\label{sec:Int}
Exotic mesons $QQ\overline{q}\overline{q}$ containing two heavy quarks $Q$
are already on focus of theoretical studies starting from pioneering works
\cite{Ader:1981db,Lipkin:1986dw,Zouzou:1986qh,Carlson:1987hh}. An important
problem analyzed in these articles was stability of four-quark states $QQ%
\overline{q}\overline{q}$ against strong decays, and processes in which they
may be discovered. It was shown that tetraquarks $QQ\overline{q}\overline{q}$
might be stable if a ratio $m_{Q}/m_{q}$ is sufficiently large. Evidently,
tetraquarks composed of a diquark $bb$ and a light antidiquark are first
candidates to stable four-quark mesons. Indeed, the isoscalar axial-vector
tetraquark $T_{bb;\overline{u}\overline{d}}^{-}$ is expected to be below the
two $B$-meson threshold and strong-interaction stable state \cite%
{Carlson:1987hh}. In last years, investigations performed using different
methods and models confirmed stable nature of the $T_{bb;\overline{u}%
\overline{d}}^{-}$ \cite%
{Navarra:2007yw,Karliner:2017qjm,Eichten:2017ffp,Agaev:2018khe}. Moreover,
it became possible to identify other $bb\overline{q}\overline{q}^{\prime }$
tetraquarks as states stable against strong and electromagnetic decays, and
calculate their full width via allowed weak transformations \cite%
{Xing:2018bqt,Agaev:2020mqq,Agaev:2020zag,Agaev:2020dba,Agaev:2019lwh}.

The status of particles $bc\overline{q}\overline{q}^{\prime }$ and $cc%
\overline{q}\overline{q}^{\prime }$ in this sense is not quite clear: they
may exist either as bound or resonant states. Nevertheless, these particles
are interesting objects for investigations, and deserve detailed analysis.
In this paper, we consider the axial-vector tetraquark $cc\overline{u}%
\overline{d}$, and therefore, in what follows, concentrate on properties of
states $cc\overline{q}\overline{q}^{\prime }$. Thus, the axial-vector
tetraquark $cc\overline{u}\overline{d}$ was studied using the QCD sum rule
method in Ref.\ \cite{Navarra:2007yw}. Exotic mesons with a general content $%
cc\overline{q}\overline{q}^{\prime }$ and quantum numbers $J^{\mathrm{P}%
}=0^{-},~0^{+},~1^{-}$ and $1^{+}$ were investigated in Ref.\ \cite%
{Du:2012wp} in the framework of the same approach. The axial-vector state $cc%
\overline{u}\overline{d}$ was modeled as a hadronic molecule composed of the
conventional mesons $D^{0}$ and $D^{\ast +}$ as well \cite%
{Dias:2011mi,Li:2012ss}.

Recent intensive analyses of heavy tetraquarks were triggered by observation
of doubly charmed baryon $\Xi _{cc}^{++}=ccu$ \cite{Aaij:2017ueg}.
Parameters of this particle were utilized as input information in a
phenomenological model to evaluate mass of the axial-vector tetraquark $\
T_{cc;\overline{u}\overline{d}}^{+}$ \cite{Karliner:2017qjm}. It was
demonstrated that $T_{cc;\overline{u}\overline{d}}^{+}$ is unstable particle
and can decay to $D^{0}D^{\ast +}$ mesons. Similar conclusions about $T_{cc;%
\overline{u}\overline{d}}^{+}$ were also made in Refs.\ \cite%
{Eichten:2017ffp,Wang:2017uld,Wang:2017dtg,Braaten:2020nwp,Cheng:2020wxa}.
Contrary, existence of stable axial-vector tetraquark was predicted in Ref.\
\cite{Meng:2020knc}, in which the authors used a constituent quark model and
found that the mass of $T_{cc;\overline{u}\overline{d}}^{+}$ is $23~\mathrm{%
MeV}$ below the two-meson threshold. In accordance with lattice simulations
the mass of the axial-vector state $cc\overline{u}\overline{d}$ is below the
two-meson threshold and this gap is equal to $(23\pm 11)~\mathrm{MeV}$ \cite%
{Junnarkar:2018twb}.

The pseudoscalar and scalar exotic mesons $cc\overline{u}\overline{d}$ were
considered in a detailed form in our paper \cite{Agaev:2019qqn}. Our studies
proved that these tetraquarks are unstable resonances, and decay strongly to
ordinary mesons. To compute widths of these particles, we utilized their
kinematically allowed decays to $D^{+}D^{\ast }(2007)^{0}$, $D^{0}D^{\ast
}(2010)^{+}$, and $D^{0}D^{+}$ meson pairs. Another subclass of doubly
charmed tetraquarks contains particles and bear two units of electric
charge. Spectroscopic parameters and widths of the such pseudoscalar states $%
cc\overline{s}\overline{s}$ and $cc\overline{d}\overline{s}$ were computed
in Ref.\ \cite{Agaev:2018vag}.

Investigation of doubly charmed exotic mesons is not limited by calculation
of their parameters using different methods such as the chiral quark model,
dynamical and relativistic quark models: Production of these particles in
ion, proton-proton and electron-positron collisions, in $B_{c}$ and $\Xi
_{bc}$ decays was investigated as well (see Ref.\ \cite{Qin:202zlg}, and
references in \cite{Agaev:2019qqn,Agaev:2018vag})\textbf{.}

First experimental information on doubly charmed exotic meson was announced
recently by the LHCb collaboration \cite{Aaij:2021vvq,LHCb:2021auc}. The
axial-vector exotic meson $T_{cc}^{+}$ with content $cc\overline{u}\overline{%
d}$ was discovered in $D^{0}D^{0}\pi ^{+}$ invariant mass distribution as a
narrow peak. Its mass is below, but very close to the $D^{0}D^{\ast
}(2010)^{+}$ threshold. The two-meson $D^{0}D^{\ast }{}^{+}$ threshold
amounts to $3875.1~\mathrm{MeV}$, whereas $T_{cc}^{+}$ has the mass%
\begin{equation}
m_{\exp }=3875.1~\mathrm{MeV}+\delta m_{\exp },  \label{eq.Exp}
\end{equation}%
where $\delta m_{\exp }$ was measured to be equal to
\begin{equation}
\delta m_{\exp }=-273\pm 61\pm 5_{-14}^{+11}~\mathrm{KeV}.  \label{eq:Delta}
\end{equation}%
The tetraquark $T_{cc}^{+}$ has the width \
\begin{equation}
\Gamma =410\pm 165\pm 43_{-38}^{+18}~\mathrm{KeV},  \label{eq:Width}
\end{equation}%
and is longest living exotic meson discovered till now.

Because $T_{cc}^{+}$ is very narrow state, its decay channels attracted
close attention of researches \cite%
{Feijoo:2021ppq,Yan:2021wdl,Fleming:2021wmk}. In these papers relevant
problems were addressed using different methods and models.

In the present article, we treat $T_{cc}^{+}$ as a doubly charmed
axial-vector diquark-antidiquark state with quark content $cc\overline{u}%
\overline{d}$ and calculate its spectroscopic parameters, as well as
evaluate width of this resonance. Calculation of the mass $m$ and current
coupling $f$ of the $T_{cc}^{+}$ are carried out by means of the QCD
two-point sum rule method, which is one of powerful nonperturbative
approaches to evaluate parameters of  conventional hadrons \cite%
{Shifman:1978bx,Shifman:1978by}. But it can be applied to extract masses and
couplings of multiquark particles \cite{Albuquerque:2018jkn}, which was
successfully demonstrated to explore numerous tetraquarks (see, for example,
Ref.\ \cite{Agaev:2020zad}).

The four-quark exotic meson $T_{cc}^{+}$ was discovered in the $%
D^{0}D^{0}\pi ^{+}$ mass distribution, and hence decays strongly to these
mesons. One of possible and most discussed ways to explain this
transformation is the chain of decays $T_{cc}^{+}\rightarrow D^{0}D^{\ast
}{}^{+}\rightarrow D^{0}D^{0}\pi ^{+}$. But the process $T_{cc}^{+}%
\rightarrow D^{0}D^{\ast }{}^{+}$ is kinematically forbidden, because the
mass of $T_{cc}^{+}$ is smaller than $D^{0}D^{\ast }{}^{+}$ threshold.
Alternatively, production of $D^{0}D^{0}\pi ^{+}$ can proceed through decay
of $T_{cc}^{+}$ to a scalar tetraquark $T_{cc;\overline{u}\overline{u}}^{0}$
and $\pi ^{+}$ followed by the process $T_{cc;\overline{u}\overline{u}%
}^{0}\rightarrow D^{0}D^{0}$. Another decay channel of $T_{cc}^{+}$ is $%
T_{cc}^{+}\rightarrow \widetilde{T}\pi ^{0}\rightarrow D^{0}D^{+}\pi ^{0}$,
where $\widetilde{T}=cc\overline{u}\overline{d}$ is again a scalar
tetraquark. In other words, we consider allowed decays of $T_{cc}^{+}$ to
scalar tetraquarks as a main mechanism for transformation of $T_{cc}^{+}$.
Partial widths of these decays to scalar tetraquarks $T_{cc;\overline{u}%
\overline{u}}^{0}$ and $\widetilde{T}$ can be employed to evaluate the full
width of $\ T_{cc}^{+}$. Of course, all these argumentations are valid only
if masses of intermediate scalar tetraquarks meet necessary kinematical
restrictions. The mass and coupling of the scalar tetraquark $\widetilde{T}$
and width of its decay to a pair of mesons $D^{0}D^{+}$ were calculated in
our work \cite{Agaev:2019qqn}. It turns out that parameters of $\widetilde{T}
$ satisfy required constraints provided one takes into account theoretical
uncertainties of the sum rule computations. In the present article, we
concentrate on the decay $T_{cc}^{+}\rightarrow \widetilde{T}\pi ^{0}$ and
evaluate its partial width. To this end, we calculate the strong coupling $g$%
, which corresponds to the vertex $T_{cc}^{+}\widetilde{T}\pi ^{0}$, and
extract its numerical value applying the QCD three-point sum rule approach.
The partial with of the process $T_{cc}^{+}\rightarrow T_{cc;\overline{u}%
\overline{u}}^{0}\pi ^{+}$ can be estimated using the isospin symmetry and a
result obtained for the first decay channel.

This paper is organized in the following way: In Sec.\ \ref{sec:Mass}, we
calculate the mass and current coupling of the tetraquark $T_{cc}^{+}$ in
the framework of the QCD two-point sum rule method by taking into account
various vacuum condensates up to dimension $10$. Section \ref{sec:Width} is
devoted to analysis of the decay channel $T_{cc}^{+}\rightarrow \widetilde{T}%
\pi ^{0}$, to calculation of the strong coupling $g$ and partial width of
this process. The width of the second channel $T_{cc}^{+}\rightarrow T_{cc;%
\overline{u}\overline{u}}^{0}\pi ^{+}$ and full width of the tetraquark $%
T_{cc}^{+}$ are evaluated in this section as well. We reserve Sec.\ \ref%
{sec:Conclusion} for discussion and concluding notes.


\section{Mass and current coupling of $T_{cc}^{+}$}

\label{sec:Mass}

We consider $T_{cc}^{+}$ as an axial-vector diquark-antidiquark state
composed of axial-vector diquark $c^{T}C\gamma _{\mu }c$ and light scalar
antidiquark $\overline{u}\gamma _{5}C\overline{d}^{T}$. The interpolating
current for such state is given by the expression

\begin{equation}
J_{\mu }(x)=c_{a}^{T}(x)C\gamma _{\mu }c_{b}(x)\overline{u}_{a}(x)\gamma
_{5}C\overline{d}_{b}^{T}(x),  \label{eq:Curr1}
\end{equation}%
where $a$ and $b$ are color indices and $C$ is charge-conjugation matrix.
The current $J_{\mu }$ belongs to the $[\overline{\mathbf{3}}%
_{c}]_{cc}\otimes \lbrack \mathbf{3}_{c}]_{\overline{u}\overline{d}}$
representation of the color group $SU_{c}(3)$, and should have lowest mass
in its class \cite{Jaffe:2004ph}.

The sum rules for the mass $m$ and current coupling $f$ of the tetraquark $%
T_{cc}^{+}$ can be obtained from analysis of the two-point correlation
function $\Pi _{\mu \nu }(p)$ given by the formula
\begin{equation}
\Pi _{\mu \nu }(p)=i\int d^{4}xe^{ipx}\langle 0|\mathcal{T}\{J_{\mu
}(x)J_{\nu }^{\dag }(0)\}|0\rangle.  \label{eq:CF1}
\end{equation}

To extract required sum rules and determine their phenomenological side, we
express the correlation function $\Pi _{\mu \nu }(p)$ in terms of the
tetraquarks' physical parameter. Because $T_{cc}^{+}$ has lowest mass in the
class of axial-vector states with the same quark content, we treat it as
ground-state particle, and keep explicitly only the first term in $\Pi _{\mu
\nu }^{\mathrm{Phys}}(p)$
\begin{equation}
\Pi _{\mu \nu }^{\mathrm{Phys}}(p)=\frac{\langle 0|J_{\mu
}|T_{cc}^{+}(p,\epsilon )\rangle \langle T_{cc}^{+}(p,\epsilon )|J_{\nu
}^{\dagger }|0\rangle }{m^{2}-p^{2}}+\cdots .  \label{eq:PhysSide}
\end{equation}%
The $\Pi _{\mu \nu }^{\mathrm{Phys}}(p)$ is derived by saturating the
correlation function (\ref{eq:CF1}) with a complete set of states with
quantum numbers $J^{\mathrm{P}}=1^{+}$ and performing the integration over $%
x $. The dots in Eq.\ (\ref{eq:PhysSide}) stand for contributions to $\Pi
_{\mu \nu }^{\mathrm{Phys}}(p)$ arising from higher resonances and continuum
states.

The function $\Pi _{\mu \nu }^{\mathrm{Phys}}(p)$ can be rewritten using the
matrix element
\begin{equation}
\langle 0|J_{\mu }|T_{cc}^{+}(p,\epsilon )\rangle =fm\epsilon _{\mu },
\label{eq:MElem1}
\end{equation}%
where $\epsilon _{\mu }$ is the polarization vector of the state $T_{cc}^{+}$%
. It is easy to show that in terms of $m$ and $f$ \ the function $\Pi _{\mu
\nu }^{\mathrm{Phys}}(p)$ takes the following form
\begin{equation}
\Pi _{\mu \nu }^{\mathrm{Phys}}(p)=\frac{m^{2}f^{2}}{m^{2}-p^{2}}\left(
-g_{\mu \nu }+\frac{p_{\mu }p_{\nu }}{m^{2}}\right) +\cdots.
\label{eq:PhysSide1}
\end{equation}

The QCD side of the sum rules $\Pi _{\mu \nu }^{\mathrm{OPE}}(p)$ should be
computed in the operator product expansion ($\mathrm{OPE}$) with certain
accuracy. To get $\Pi _{\mu \nu }^{\mathrm{OPE}}(p)$, we insert into Eq.\ (%
\ref{eq:CF1}) the interpolating current $J_{\mu }(x)$, and contract relevant
heavy and light quark fields. After these manipulations, we find
\begin{eqnarray}
&&\Pi _{\mu \nu }^{\mathrm{OPE}}(p)=i\int d^{4}xe^{ip\cdot x}\left\{ \mathrm{%
Tr}\left[ \gamma _{5}\widetilde{S}_{d}^{b^{\prime }b}(-x)\gamma
_{5}S_{u}^{a^{\prime }a}(-x)\right] \right.  \notag \\
&&\times \mathrm{Tr}\left[ \gamma _{\nu }\widetilde{S}_{c}^{aa^{\prime
}}(x)\gamma _{\mu }S_{c}^{bb^{\prime }}(x)\right] -\mathrm{Tr}\left[ \gamma
_{5}\widetilde{S}_{d}^{b^{\prime }b}(-x)\right.  \notag \\
&&\left. \left. \times \gamma _{5}S_{u}^{a^{\prime }a}(-x)\right] \mathrm{Tr}%
\left[ \gamma _{\nu }\widetilde{S}_{c}^{ba^{\prime }}(x)\gamma _{\mu
}S_{c}^{ab^{\prime }}(x)\right] \right\}.  \label{eq:QCDSide}
\end{eqnarray}%
In Eq.\ (\ref{eq:QCDSide}) $S_{c}^{ab}(x)$ and $S_{q}^{ab}(x)$ are the $c$
and $q(u,d)$-quark propagators: their explicit expressions can be found in
Ref.\ \cite{Agaev:2020zad}. Here, we also use the short-hand notation
\begin{equation}
\widetilde{S}_{c(q)}(x)=CS_{c(q)}^{T}(x)C.  \label{eq:Prop}
\end{equation}

The QCD sum rules can be extracted by employing the same Lorentz structures
both in $\Pi _{\mu \nu }^{\mathrm{Phys}}(p)$ and $\Pi _{\mu \nu }^{\mathrm{%
OPE}}(p)$. The structures proportional to $g_{\mu \nu }$ are appropriate for
our purposes, because they receive contributions only from spin-$1$
particles. We denote corresponding invariant amplitudes by $\Pi ^{\mathrm{%
Phys}}(p^{2})$ and $\Pi ^{\mathrm{OPE}}(p^{2})$, respectively. Another
problem to be solved is suppression of contributions coming from the higher
resonances and continuum states. To this end, one should apply the Borel
transformation to both sides of the sum rule equality. At the next stage
using the quark-hadron duality hypothesis, one has to subtract higher
resonance and continuum terms from the physical side of the equality. As a
result, the sum rule equality depends on the Borel $M^{2}$ and continuum
threshold $s_{0}$ parameters.

The Borel transformation of $\Pi ^{\mathrm{Phys}}(p^{2})$ is trivial. The
Borel transformed and continuum subtracted invariant amplitude $\Pi ^{%
\mathrm{OPE}}(p^{2})$ has rather complicated form
\begin{equation}
\Pi (M^{2},s_{0})=\int_{4m_{c}^{2}}^{s_{0}}ds\rho ^{\mathrm{OPE}%
}(s)e^{-s/M^{2}}+\Pi (M^{2}).  \label{eq:InvAmp}
\end{equation}%
Here, $\rho ^{\mathrm{OPE}}(s)$ is the two-point spectral density, whereas
second component of the invariant amplitude $\Pi (M^{2})$ includes
nonperturbative contributions calculated directly from $\Pi _{\mu \nu }^{%
\mathrm{OPE}}(p)$. We compute $\Pi (M^{2},s_{0})$ by taking into account
nonperturbative terms up to dimension $10$.

Then, the sum rules for $m$ and $f$ read
\begin{equation}
m^{2}=\frac{\Pi ^{\prime }(M^{2},s_{0})}{\Pi (M^{2},s_{0})},  \label{eq:Mass}
\end{equation}%
and
\begin{equation}
f^{2}=\frac{e^{m^{2}/M^{2}}}{m^{2}}\Pi (M^{2},s_{0}),  \label{eq:Coupling}
\end{equation}%
where $\Pi ^{\prime }(M^{2},s_{0})=d/d(-1/M^{2})\Pi (M^{2},s_{0})$.

The expressions (\ref{eq:Mass}) and (\ref{eq:Coupling}) contain various
quark, gluon and mixed condensates, which are universal parameters:
\begin{eqnarray}
&&\langle \overline{q}q\rangle =-(0.24\pm 0.01)^{3}~\mathrm{GeV}^{3},\
\notag \\
&&\langle \overline{q}g_{s}\sigma Gq\rangle =m_{0}^{2}\langle \overline{q}%
q\rangle ,\ m_{0}^{2}=(0.8\pm 0.1)~\mathrm{GeV}^{2},\   \notag \\
&&\langle \frac{\alpha _{s}G^{2}}{\pi }\rangle =(0.012\pm 0.004)~\mathrm{GeV}%
^{4},  \notag \\
&&\langle g_{s}^{3}G^{3}\rangle =(0.57\pm 0.29)~\mathrm{GeV}^{6},  \notag \\
&&m_{c}=1.275\pm 0.025~\mathrm{GeV}.  \label{eq:Parameters}
\end{eqnarray}%
The mass of the $c$\ quark is also included into this list. The Borel and
continuum threshold parameters $M^{2}$ and $s_{0}$ are auxiliary quantities
of computations and their choice should satisfy constraints imposed on the
pole contribution ($\mathrm{PC}$) and convergence of $\mathrm{OPE}$. A
minimum sensitivity of the extracted quantities to the Borel parameter $%
M^{2} $ is also among important requirements. The maximum value of $M^{2}$
can be obtained from the restriction on $\mathrm{PC}$%
\begin{equation}
\mathrm{PC}=\frac{\Pi (M^{2},s_{0})}{\Pi (M^{2},\infty )}.  \label{eq:Pole}
\end{equation}%
In the present work, we apply the constraint $\mathrm{PC\geqslant 0.2}$,
which is usual for the multiquark hadrons. The low limit of the working
region for the Borel parameter is fixed from convergence of the operator
product expansion, i.e., from analysis of the ratio%
\begin{equation}
R(M^{2})=\frac{\Pi ^{\mathrm{DimN}}(M^{2},s_{0})}{\Pi (M^{2},s_{0})},
\label{eq:Convergence}
\end{equation}%
where $\Pi ^{\mathrm{DimN}}(M^{2},s_{0})$ denotes contribution of the last
three terms in $\mathrm{OPE}$, in other words $\mathrm{DimN=Dim(8+9+10)}$.
The convergence of the operator product expansion at minimum of $M^{2}$ is
ensured by the requirement $R(M^{2})\leq 0.01$.

Our analysis demonstrates that working regions for $M^{2}$ and $s_{0}$
\begin{equation}
M^{2}\in \lbrack 4,6]~\mathrm{GeV}^{2},\ s_{0}\in \lbrack 19.5,21.5]~\mathrm{%
GeV}^{2},  \label{eq:Regions}
\end{equation}%
meet all aforementioned constraints. Thus, in these regions $\mathrm{PC}$
changes on average within limits
\begin{equation}
0.62\leq \mathrm{PC}\leq 0.20.  \label{eq:Polelimits}
\end{equation}%
At the minimum $M^{2}=4~\mathrm{GeV}^{2}$ contributions to $\Pi
(M^{2},s_{0}) $ coming from last three terms in $\mathrm{OPE}$ do not exceed
$1\%$ of the full result.

Central values of the mass $m$ and coupling $f$ are evaluated by averaging
results for these parameters over working regions (\ref{eq:Regions}).
Obtained mean values correspond to predictions of the sum rules
approximately at middle point of these regions, i.e., to results at $M^{2}=5~%
\mathrm{GeV}^{2}$ and $s_{0}=20.4~\mathrm{GeV}^{2}$. At this point the pole
contribution is $\mathrm{PC}\approx 0.55$, which guarantees the ground-state
nature of $T_{cc}^{+}$. Our results for $m$\ and $f$\ are
\begin{eqnarray}
m &=&(3868~\pm 124)~\mathrm{MeV},  \notag \\
f &=&(5.03\pm 0.79)\times 10^{-3}~\mathrm{GeV}^{4}.  \label{eq:Result1}
\end{eqnarray}

\begin{figure}[h]
\includegraphics[width=8.8cm]{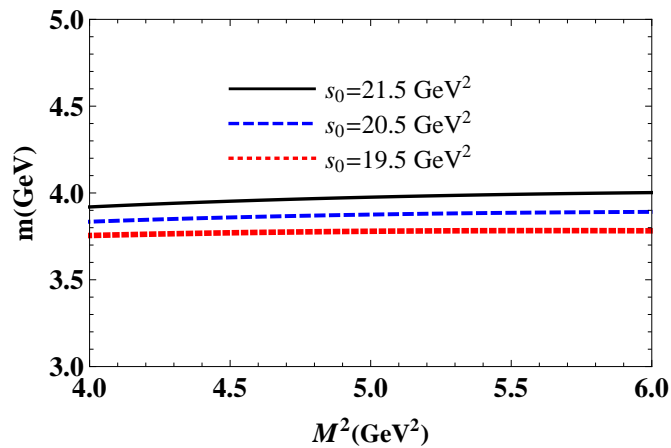}
\caption{The mass of the tetraquark $T^{+}_{cc}$ as a function of the Borel
parameter $M^2$ at fixed $s_0$.}
\label{fig:Mass}
\end{figure}

In general, quantities extracted from the sum rules should not depend on the
choice of the parameter $M^{2}$. But, in real calculations, one observes a
residual dependence of $m$ and $f$ on $M^{2}$. In Fig.\ \ref{fig:Mass}, we
depict the mass $m$ of the tetraquark as a function of $M^{2}$. It is seen,
that the region for $M^{2}$ shown in this figure can be considered as a
relatively stable plateau, where parameters of $T_{cc}^{+}$ can be
evaluated. This dependence on $M^{2}$ allows us also to estimate
uncertainties generated by the sum rule calculations. The second source of
theoretical errors is a choice of the continuum threshold parameter $s_{0}$.
The working window for $s_{0}$ should satisfy limits arising from dominance
of $\mathrm{PC}$ and convergence of $\mathrm{OPE}$ as well. Additionally, $%
s_{0}$ carries physical information about first excitation of the tetraquark
$T_{cc}^{+}$. The self-consistent analysis implies that $\sqrt{s_{0}}$ is
smaller than mass of such state. In the case under discussion, the mass gap
is $\sqrt{s_{0}}-m\approx 650~\mathrm{MeV}$ which can be considered as a
reasonable estimate for exotic mesons containing two heavy $c$ quarks.

Effects connected with a choice of parameters $M^{2}$ and $s_{0}$ are two
main sources of theoretical uncertainties in sum rule computations. In the
case of the mass $m$ they equal to $\pm 3.2\%$, whereas for the coupling $f$
ambiguities are $\pm 16\%$ of the full result. Theoretical uncertainties for
$f$ are larger than that for the mass, but they do not exceed accepted
limits.

The central value of the mass $m$ extracted from the sum rules is below the
two-meson $D^{0}D^{\ast }(2010)^{+}$ threshold. But a mass gap $\delta m$
between $m=3868\ \mathrm{MeV}$ and two-meson threshold is $\sim 7\ \mathrm{%
MeV}$ and considerably overshoots experimental value. It is known, that the
sum rule method suffers from uncertainties which do not allow one to improve
accuracy of performed analysis. Principal result of the present
study is prediction for the mass of the exotic meson $T_{cc}^{+}$ which is
below the two-meson threshold and, in this respect, compatible with LHCb
observation.

The mass and coupling of the $T_{cc}^{+}$\ have been obtained from the sum
rules using for their physical side $\Pi _{\mu \nu }^{\mathrm{Phys}}(p)$\ a
simple-pole approximation [see, Eq.\ (\ref{eq:PhysSide})]. In the case of
the multiquark hadrons such approximation should be justified by additional
arguments, because a physical side of relevant sum rules receives
contributions from two-hadron reducible terms as well. This problem was
first posed during theoretical studies of the pentaquarks \cite%
{Kondo:2004cr,Lee:2004xk}. Two-hadron contaminating terms have to be taken
into account when extracting parameters of multiquark hadrons. In the case
of the tetraquarks they lead to modification in the quark propagator \cite%
{Wang:2015nwa}%
\begin{equation}
\frac{1}{m^{2}-p^{2}}\rightarrow \frac{1}{m^{2}-p^{2}-i\sqrt{p^{2}}\Gamma (p)%
},  \label{eq:Modif}
\end{equation}%
\textbf{w}here $\Gamma (p)$\ is the finite width of the tetraquark generated
by two-meson states. Investigations demonstrate that these effects rescale
the original coupling $f$\ and leave stable the mass $m$\ of the tetraquark
\cite{Agaev:2018vag,Sundu:2018nxt}. Detailed analyses proved that two-meson
contributions are small even for tetraquarks with width around a hundred $%
\mathrm{MeV}$\ \cite{Agaev:2018vag,Sundu:2018nxt}. In the case under
discussion, the width of the $T_{cc}^{+}$\ is less than one $\mathrm{MeV}$,
therefore two-meson effects can be safely neglected.


\section{Width of the tetraquark $T_{cc}^{+}$}

\label{sec:Width}

As was noted above, $T_{cc}^{+}$ was discovered in $D^{0}D^{0}\pi ^{+}$ mass
distribution, and hence, can decay strongly to these mesons. This process
may run through $T_{cc}^{+}\rightarrow D^{0}D^{\ast +}$ followed by the
decay $D^{\ast +}\rightarrow D^{0}\pi ^{+}$. The experimental measurements
showed, however, that the mass of $T_{cc}^{+}$ is not enough to trigger this
mechanism. In this situation, the final state $D^{0}D^{0}\pi ^{+}$ can be
achieved due to production of the intermediate scalar tetraquark $T_{cc;%
\overline{u}\overline{u}}^{0}$. But the $T_{cc}^{+}$ may decay to mesons $%
D^{0}D^{+}\pi ^{0}$, as well.

In this section, we consider the decay $T_{cc}^{+}\rightarrow \widetilde{T}%
\pi ^{0}$, and calculate its partial width. The reason is that parameters of
the tetraquark $\widetilde{T}$ are known and were calculated in our paper
\cite{Agaev:2019qqn}%
\begin{eqnarray}
m_{\widetilde{T}} &=&(3845~\pm 175)~\mathrm{MeV},  \notag \\
f_{\widetilde{T}} &=&(1.16\pm 0.26)\times 10^{-2}~\mathrm{GeV}^{4}.
\label{eq:Results2}
\end{eqnarray}%
To realize the process $T_{cc}^{+}\rightarrow \widetilde{T}\pi
^{0}\rightarrow D^{0}D^{+}\pi ^{0}$ the mass of $\widetilde{T}$ should
satisfy the constraints $m_{D^{0}D^{+}}<m_{\widetilde{T}}$ $<m-m_{\pi }$. In
other words, $\widetilde{T}$ has to be heavier than $m_{D^{0}D^{+}}\approx
3735~\mathrm{MeV}$. In this section, we use for the mass of the $T_{cc}^{+}$
experimental value $m\approx 3875~\mathrm{MeV},$ therefore $\widetilde{T}$
should be lighter than $3740~\mathrm{MeV}$. It is seen that the narrow
region allowed for the mass of the tetraquark $\widetilde{T}$ between $3735~%
\mathrm{MeV}<m_{\widetilde{T}}<3740~\mathrm{MeV}$ is consistent with
prediction for $m_{\widetilde{T}}$ from Eq.\ (\ref{eq:Results2}). For our
computations, we fix $m_{\widetilde{T}}$ $=3736~\mathrm{MeV}$ and use it in
what follows. Accordingly, \ for the current coupling, we employ\textbf{\ }$%
f_{\widetilde{T}}=1.42\times 10^{-2}~\mathrm{GeV}^{4}$.

Here, some comments concerning strong decay channels of $\widetilde{T}$ are
necessary. They were analyzed in Ref.\ \cite{Agaev:2019qqn}, in which it was
demonstrated that the exotic state $\widetilde{T}$ in $S$-wave decays to a
pair of the mesons $D^{0}D^{+}$, but its $P$-wave channels require a master
particle $\widetilde{T}$ to be considerably heavier than $3736~\mathrm{MeV}$%
, which is not the case. Stated differently, only open strong decay channel
of $\widetilde{T}$ is the process $\widetilde{T}\rightarrow D^{0}D^{+}$.
Hence, width of the tetraquark $\widetilde{T}$ is determined by this decay
and equals to
\begin{equation}
\Gamma \lbrack \widetilde{T}\rightarrow D^{0}D^{+}]=(12.4\pm 3.1)~\mathrm{MeV%
}.  \label{eq:DW2a}
\end{equation}

Below we investigate the decay $T_{cc}^{+}\rightarrow \widetilde{T}\pi ^{0}$
and calculate the strong coupling corresponding to the vertex $T_{cc}^{+}%
\widetilde{T}\pi ^{0}$. To derive the QCD three-point sum rule for this
coupling and extract its numerical value, we start from analysis of the
correlation function
\begin{eqnarray}
&&\Pi _{\mu }(p,p^{\prime })=i^{2}\int d^{4}xd^{4}ye^{i(p^{\prime
}y-px)}\langle 0|\mathcal{T}\{\widetilde{J}(y)  \notag \\
&&\times J_{\pi }(0)J_{\mu }^{\dagger }(x)\}|0\rangle,  \label{eq:CF2}
\end{eqnarray}%
where $J_{\mu }(x)$,$\ \widetilde{J}(x)$ and $J_{\pi }(x)$ are the
interpolating currents for the tetraquarks $T_{cc}^{+}$ and $\widetilde{T}$
\ and the pion $\pi ^{0}$, respectively. The $J_{\mu }(x)$ is given by Eq.\ (%
\ref{eq:Curr1}), for two remaining currents, we employ
\begin{eqnarray}
&&\widetilde{J}(x)=\epsilon \widetilde{\epsilon }[c_{b}^{T}(x)C\gamma _{\mu
}c_{c}(x)][\overline{u}_{d}(x)\gamma ^{\mu }C\overline{d}_{e}^{T}(x)],\ \
\notag \\
&&J_{\pi }(x)=\frac{1}{\sqrt{2}}\left[ \overline{u}_{i}(x)i\gamma
_{5}u_{i}(x)-\overline{d}_{i}(x)i\gamma _{5}d_{i}(x)\right].
\label{eq:Curr5}
\end{eqnarray}%
Here, $\epsilon \widetilde{\epsilon }=\epsilon _{abc}\epsilon _{ade}$, and $%
a,b,c,d$, $e$ and $i$ are color indices. The 4-momenta of the tetraquarks $%
T_{cc}^{+}$ and $\widetilde{T}$ are denoted by $p$ and $p^{\prime }$, as a
result, the momentum of the pion $\pi ^{0}$ is $q=p-p^{\prime }$.

We follow the usual recipes of the sum rule method and, first calculate the
correlation function $\Pi _{\mu }(p,p^{\prime })$ using phenomenological
parameters of the involved particles. Separating the ground-state
contribution to the correlation function (\ref{eq:CF2}) from effects of
higher resonances and continuum states,\ for the physical side of the sum
rule $\Pi _{\mu }^{\mathrm{Phys}}(p,p^{\prime })$, we obtain%
\begin{eqnarray}
&&\Pi _{\mu }^{\mathrm{Phys}}(p,p^{\prime })=\frac{\langle 0|\widetilde{J}|%
\widetilde{T}(p^{\prime })\rangle \langle 0|J_{\pi }|\pi ^{0}(q)\rangle }{%
(p^{\prime 2}-m_{\widetilde{T}}^{2})(q^{2}-m_{\pi }^{2})}  \notag \\
&&\times \frac{\langle T_{cc}^{+}(p,\epsilon )|J_\mu^{\dagger }|0\rangle
\langle \pi ^{0}(q)\widetilde{T}(p^{\prime })|T_{cc}^{+}(p,\epsilon )\rangle
}{(p^{2}-m^{2})}+\cdots,  \label{eq:CF3}
\end{eqnarray}%
with $m_{\pi }$ being the mass of the pion.

To simplify further the function $\Pi _{\mu }^{\mathrm{Phys}}(p,p^{\prime })$%
, it is convenient to use matrix elements of the tetraquarks and pion. To
this end, we introduce the matrix elements
\begin{eqnarray}
&&\langle 0|\widetilde{J}|\widetilde{T}(p^{\prime })\rangle =m_{\widetilde{T}%
}f_{\widetilde{T}},\   \notag \\
&&\langle 0|\overline{u}i\gamma _{5}u|\pi ^{0}(q)\rangle =\frac{1}{\sqrt{2}}%
f_{\pi }\mu _{\pi },\ \mu _{\pi }=-\frac{2\langle \overline{q}q\rangle }{%
f_{\pi }^{2}},  \label{eq:Mel2}
\end{eqnarray}%
where $f_{\pi }$ and $\langle \overline{q}q\rangle $ are the pion decay
constant and the quark vacuum condensate, respectively. The matrix element
of the $d$ quark field $\overline{d}i\gamma _{5}d$ is given by the similar
expression.

We model $\langle \pi (q)\widetilde{T}(p^{\prime })|T_{cc}^{+}(p,\epsilon
)\rangle $ in the form%
\begin{equation}
\langle \pi (q)\widetilde{T}(p^{\prime })|T_{cc}^{+}(p,\epsilon )\rangle
=g(q^{2})p_{\mu }^{\prime }\epsilon ^{\ast \mu }  \label{eq:Ver1}
\end{equation}%
and denote by $g(q^{2})$ the strong coupling at the vertex $T_{cc}^{+}%
\widetilde{T}\pi ^{0}$. Then, it is easy to find that
\begin{eqnarray}
&&\Pi _{\mu }^{\mathrm{Phys}}(p,p^{\prime })=g(q^{2})i\frac{f_{\pi }\mu
_{\pi }f_{\widetilde{T}}m_{\widetilde{T}}fm}{(p^{\prime 2}-m_{\widetilde{T}%
}^{2})(q^{2}-m_{\pi }^{2})}  \notag \\
&&\times \frac{1}{(p^{2}-m^{2})}\left( \frac{m^{2}+m_{\widetilde{T}%
}^{2}-q^{2}}{2m^{2}}p_{\mu }-p_{\mu }^{\prime }\right) +\cdots.
\label{eq:Phys2}
\end{eqnarray}%
The double Borel transformation of the correlation function over variables $%
p^{2}$ and $p^{\prime 2}$ is given by the following formula
\begin{eqnarray}
&&\mathcal{B}\Pi _{\mu }^{\mathrm{Phys}}(p,p^{\prime })=g(q^{2})i\frac{%
f_{\pi }\mu _{\pi }f_{\widetilde{T}}m_{\widetilde{T}}fm}{(q^{2}-m_{\pi }^{2})%
}e^{-m^{2}/M_{1}^{2}}  \notag \\
&&\times e^{-m_{\widetilde{T}}^{2}/M_{2}^{2}}\left( \frac{m^{2}+m_{%
\widetilde{T}}^{2}-q^{2}}{2m^{2}}p_{\mu }-p_{\mu }^{\prime }\right) +\cdots.
\end{eqnarray}%
The function $\mathcal{B}\Pi _{\mu }^{\mathrm{Phys}}(p,p^{\prime })$
contains Lorentz structures proportional to $p_{\mu }$ and $p_{\mu }^{\prime
}$. We work with the invariant amplitude $\Pi ^{\mathrm{Phys}%
}(p^{2},p^{\prime 2},q^{2})$ corresponding to the structure proportional to $%
p_{\mu }$. The Borel transform of this amplitude forms the phenomenological
side of the sum rule.

To find the QCD side of the three-point sum rule, we compute $\Pi _{\mu
}(p,p^{\prime })$ in terms of the quark propagators and get
\begin{eqnarray}
&&\Pi _{\mu }^{\mathrm{OPE}}(p,p^{\prime })=i^{3}\sqrt{2}\int
d^{4}xd^{4}y\epsilon ^{\prime }\widetilde{\epsilon }^{\prime }e^{i(p^{\prime
}y-px)}  \notag \\
&&\times \mathrm{Tr}\left[ \gamma _{5}S_{u}^{id^{\prime }}(-y)\gamma ^{\nu }%
\widetilde{S}_{d}^{be^{\prime }}(x-y)\gamma _{5}S_{u}^{ai}(x)\right]  \notag
\\
&&\times \left\{ \mathrm{Tr}\left[ \gamma _{\mu }\widetilde{S}%
_{c}^{b^{\prime }a}(y-x)\gamma _{\nu }S_{c}^{c^{\prime }b}(y-x)\right]
\right.  \notag \\
&&\left. -\mathrm{Tr}\left[ \gamma _{\mu }\widetilde{S}_{c}^{c^{\prime
}a}(y-x)\gamma _{\nu }S_{c}^{bb^{\prime }}(y-x)\right] \right\},
\label{eq:CF4}
\end{eqnarray}%
where $\epsilon ^{\prime }\widetilde{\epsilon }^{\prime }=\epsilon
_{nb^{\prime }c^{\prime }}\epsilon _{nd^{\prime }e^{\prime }}$. In deriving
of Eq.\ (\ref{eq:CF4}), we have taken into account that in the chiral limit $%
m_{u}=m_{d}$ adopted in the present article, both components of the pion
interpolating current give the same results: In Eq.\ (\ref{eq:CF4}), $\Pi
_{\mu }^{\mathrm{OPE}}(p,p^{\prime })$ is twice of the $\overline{u}i\gamma
_{5}u$ contribution.

The correlation function $\Pi _{\mu }^{\mathrm{OPE}}(p,p^{\prime })$ is
calculated with dimension-$6$ accuracy, and has the same Lorentz structures
as $\Pi _{\mu }^{\mathrm{Phys}}(p,p^{\prime })$. The double Borel
transformation $\mathcal{B}\Pi ^{\mathrm{OPE}}(p^{2},p^{\prime 2},q^{2})$,
where $\Pi ^{\mathrm{OPE}}(p^{2},p^{\prime 2},q^{2})$ is the invariant
amplitude that corresponds to the term proportional to $p_{\mu }$,
constitutes the second part of the sum rule. By equating $\mathcal{B}\Pi ^{%
\mathrm{OPE}}(p^{2},p^{\prime 2},q^{2})$ and Borel transformation of $\Pi ^{%
\mathrm{Phys}}(p^{2},p^{\prime 2},q^{2})$, and performing continuum
subtraction we find the sum rule for the coupling $g(q^{2})$.

The Borel transformed and subtracted amplitude $\Pi ^{\mathrm{OPE}%
}(p^{2},p^{\prime 2},q^{2})$ can be expressed using the spectral density $%
\widetilde{\rho }(s,s^{\prime },q^{2})$ which is proportional to a relevant
imaginary part of $\Pi _{\mu }^{\mathrm{OPE}}(p,p^{\prime })$
\begin{eqnarray}
&&\Pi (\mathbf{M}^{2},\mathbf{s}_{0},q^{2})=\int_{4m_{c}^{2}}^{s_{0}}ds%
\int_{4m_{c}^{2}}^{s_{0}^{\prime }}ds^{\prime }\widetilde{\rho }(s,s^{\prime
},q^{2})  \notag \\
&&\times e^{-s/M_{1}^{2}}e^{-s^{\prime }/M_{2}^{2}},  \label{eq:SCoupl}
\end{eqnarray}%
where $\mathbf{M}^{2}=(M_{1}^{2},\ M_{2}^{2})$ and $\mathbf{s}_{0}=(s_{0},\
s_{0}^{\prime })$ are the Borel and continuum threshold parameters,
respectively. Then, the sum rule for $g(q^{2})$ reads
\begin{eqnarray}
&&g(q^{2})=\frac{2m^{2}}{f_{\pi }\mu _{\pi }f_{\widetilde{T}}m_{\widetilde{T}%
}fm}\frac{q^{2}-m_{\pi }^{2}}{m^{2}+m_{\widetilde{T}}^{2}-q^{2}}  \notag \\
&&\times e^{m^{2}/M_{1}^{2}}e^{m_{\widetilde{T}}^{2}/M_{2}^{2}}\Pi (\mathbf{M%
}^{2},\mathbf{s}_{0},q^{2}).  \label{eq:SRCoup}
\end{eqnarray}%
The coupling $g(q^{2})$ is a function of $q^{2}$ and depends on the Borel
and continuum threshold parameters, which are not shown in Eq.\ (\ref%
{eq:SRCoup}) as arguments of $g$. We also introduce a new variable $%
Q^{2}=-q^{2}$ and denote the obtained function as $g(Q^{2})$.

The sum rule Eq.\ (\ref{eq:SRCoup}) contains masses and couplings of the
tetraquarks $T_{cc}^{+}$ and $\widetilde{T}$, as well as the mass and decay
constant of the pion $\pi ^{0}$. The spectroscopic parameters of the $%
T_{cc}^{+}$ have been calculated in the present work and presented in Eq.\ (%
\ref{eq:Result1}). As the mass and coupling of the tetraquark $\widetilde{T}$
, we use $m_{\widetilde{T}}$ $=3736~\mathrm{MeV}$ and $f_{\widetilde{T}%
}=1.42\times 10^{-2}\ \mathrm{GeV}^{4}$, respectively. For the mass and
decay constant of the pion, we employ the values: $m_{\pi }$ $=134.98~%
\mathrm{MeV}$ and $f_{\pi }=131~\mathrm{MeV}$.

Apart from these spectroscopic parameters, for numerical analysis of $%
g(Q^{2})$ one also needs to fix $\mathbf{M}^{2}$ and $\mathbf{s}_{0}$. The
constraints imposed on these auxiliary parameters are usual for sum rule
computations and have been discussed above. The regions for $M_{1}^{2}$ and $%
s_{0}$ correspond to the $T_{cc}^{+}$ channel, and coincide with the working
windows for these parameters determined in the mass calculations [see, Eq.\ (%
\ref{eq:Regions})]. The pair of parameters $(M_{2}^{2},\ s_{0}^{\prime })$
for the $\widetilde{T}$ channel are chosen within the limits
\begin{equation}
M_{2}^{2}\in \lbrack 4,\ 5]~\mathrm{GeV}^{2},\ s_{0}^{\prime }\in \lbrack
19,\ 20]~\mathrm{GeV}^{2}.  \label{eq:Wind3}
\end{equation}%
The extracted strong coupling $g(Q^{2})$ depends on $\mathbf{M}^{2}$ and $%
\mathbf{s}_{0}$: the working intervals for these parameters are chosen in
such a way that to minimize these uncertainties.

\begin{figure}[h]
\includegraphics[width=8.5cm]{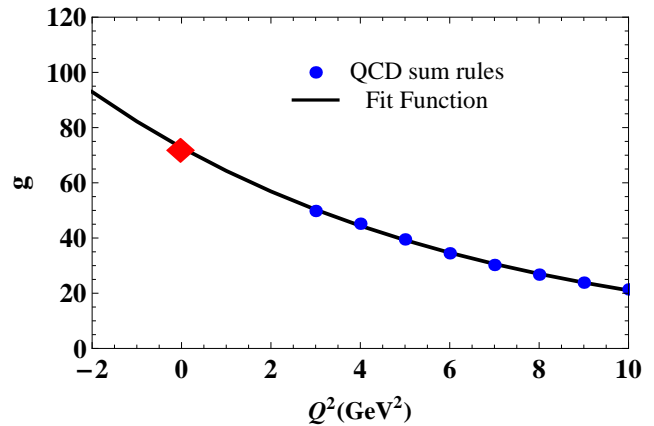}
\caption{The sum rule predictions and fit function for the strong coupling $%
g(Q^{2})$. The red diamond shows the point $Q^{2}=-m_{\protect\pi }^{2}$. }
\label{fig:Fit}
\end{figure}

The width of the decay under analysis should be computed using the strong
coupling at the pion's mass shell $q^{2}=m_{\pi }^{2}$, which is not
accessible to the sum rule calculations. We solve this problem by
introducing a fit function $G(Q^{2})$ that for the momenta $Q^{2}>0$
coincides with QCD sum rule's predictions, but can be extrapolated to the
region of $Q^{2}<0$ to find $g(-m_{\pi }^{2})$. To construct the function $%
G(Q^{2})$, we employ the analytic form
\begin{equation}
G(Q^{2})=G_{0}\mathrm{\exp }\left[ c_{1}\frac{Q^{2}}{m^{2}}+c_{2}\left(
\frac{Q^{2}}{m^{2}}\right) ^{2}\right] ,  \label{eq:FitF}
\end{equation}%
where $G_{0}$, $c_{1}$ and $c_{2}$ are fitting parameters. Numerical
analysis allows us to fix $G_{0}=72.75$\textbf{, }$c_{1}=-1.84$, and%
$c_{2}=-0.03$. In Fig.\ \ref{fig:Fit} we depict the sum
rule predictions for $g(Q^{2})$ and also provide $G(Q^{2})$: an agreement
between them is evident.

At the pion mass shell $Q^{2}=-m_{\pi }^{2}$ this function leads to
prediction
\begin{equation}
g\equiv G(-m_{\pi }^{2})=73\pm 11.  \label{eq:Coupl1}
\end{equation}%
The width of decay $T_{cc}^{+}\rightarrow \widetilde{T}\pi ^{0}$ is given by
the simple expression
\begin{equation}
\Gamma \left[ T_{cc}^{+}\rightarrow \widetilde{T}\pi ^{0}\right] =\frac{%
g^{2}\lambda ^{3}\left( m,m_{\widetilde{T}},m_{\pi }\right) }{24\pi m^{2}},
\label{eq:DW1a}
\end{equation}%
where%
\begin{equation}
\lambda \left( a,b,c\right) =\frac{1}{2a}\sqrt{a^{4}+b^{4}+c^{4}-2\left(
a^{2}b^{2}+a^{2}c^{2}+b^{2}c^{2}\right) }.
\end{equation}%
Using the strong coupling from Eq.\ (\ref{eq:Coupl1}), it is not difficult
to evaluate width of the decay $T_{cc}^{+}\rightarrow \widetilde{T}\pi ^{0}$%
\begin{equation}
\Gamma \left[ T_{cc}^{+}\rightarrow \widetilde{T}\pi ^{0}\right] =(163\pm
41)~\mathrm{keV}.  \label{eq:DW1Numeric}
\end{equation}

The second process $T_{cc}^{+}\rightarrow T_{cc;\overline{u}\overline{u}%
}^{0}\pi ^{+}$ can be considered via the same manner. Due to isospin
symmetry, the strong coupling $\widetilde{g}$ corresponding to the vertex $%
T_{cc}^{+}T_{cc;\overline{u}\overline{u}}^{0}\pi ^{+}$ is connected with $g$
through the simple relation \cite{Belyaev:1994zk}
\begin{equation}
|\widetilde{g}|=\sqrt{2}g.
\end{equation}%
As a result, the width of the second decay channel of the state $T_{cc}^{+}$
is given approximately by the expression
\begin{equation}
\Gamma \left[ T_{cc}^{+}\rightarrow T_{cc;\overline{u}\overline{u}}^{0}\pi
^{+}\right] \approx (326\pm 82)~\mathrm{keV}.  \label{eq:DW2Numeric}
\end{equation}%
Then for the full width of the exotic meson $T_{cc}^{+}$, we get
\begin{equation}
\Gamma =(489\pm 92)~\mathrm{keV},  \label{eq:FullW}
\end{equation}
in a nice agreement with the result of the LHCb collaboration.


\section{Discussion and concluding notes}

\label{sec:Conclusion}
\begin{table}[t]
\begin{tabular}{|c|c|}
\hline\hline
Works & $m$ or $\delta m$ (in units of $\mathrm{MeV}$ ) \\ \hline
This work & $3868 \pm 124$ \\
F.~S.~Navarra \textit{et al.} \cite{Navarra:2007yw} & $4000 \pm 200$ \\
J.~M.~Dias \textit{et al.} \cite{Dias:2011mi} & $3872.2 \pm 39.5$ \\
M.~Karliner, and J.~L.~Rosner \cite{Karliner:2017qjm} & $3882 \pm 12$ \\
E.~J.~Eichten, and C.~Quigg \cite{Eichten:2017ffp} & $3978$ \\
Z.~G.~Wang, and Z.~H.~Yan \cite{Wang:2017dtg} & $3900 \pm 90$ \\
E.~Braaten \textit{et al.} \cite{Braaten:2020nwp} & $3947 \pm 11$ \\
J.~B~Cheng \textit{et al.} \cite{Cheng:2020wxa} & $3929.3$ \\
Q.~Meng \textit{et al.} \cite{Meng:2020knc} & $\delta m=-23$ \\
P.~Junnarkar \textit{et al.} \cite{Junnarkar:2018twb} & $\delta m=-23 \pm 11$
\\ \hline\hline
\end{tabular}%
\caption{Theoretical predictions for the mass (or for a mass gap $\protect%
\delta m$ from the two-meson threshold) of the axial-vector state $%
T_{cc}^{+} $ obtained using different models and methods.}
\label{tab:Theory}
\end{table}

In the present work, we have calculated spectroscopic parameters and width
of the doubly charmed state $T_{cc}^{+}$ observed recently by the LHCb
collaboration. Properties of this state were studied in the context of
various methods and presented in numerous publications. Relevant information
is collected in Table\ \ref{tab:Theory}. Our result for the mass gap (for
all predictions, we consider only their central values) $\delta m=-7\
\mathrm{MeV}$ is qualitatively comparable with predictions of Refs. \cite%
{Meng:2020knc,Junnarkar:2018twb}. The table contains also prediction for $m$
found in the hadronic molecule model \cite{Dias:2011mi}, which is very close
to experimental datum. All other results are above the two-meson $%
D^{0}D^{\ast +}$ threshold, which are excluded by experimental data.

We\ have investigated strong decay channels $T_{cc}^{+}\rightarrow
\widetilde{T}\pi ^{0}$ and $T_{cc}^{+}\rightarrow T_{cc;\overline{u}%
\overline{u}}^{0}\pi ^{+}$ and by this way calculated full width of the
state $T_{cc}^{+}$. Calculation of the partial width of the process $%
T_{cc}^{+}\rightarrow \widetilde{T}\pi ^{0}$ has been carried out in the
framework of the QCD three-point sum rule method. The partial width of the
second channel has been evaluated using isospin symmetry and our result for
the first process. A very nice agreement of the $T_{cc}^{+}$ tetraquark's
full width obtained in the present article with the LHCb data considerably
strengthen arguments in favor of the diquark-antidiquark nature of the $%
T_{cc}^{+}$.

But there are still open problems that should be clarified to make firm
conclusions about inner structure of the $T_{cc}^{+}$. Main question to be
addressed is a molecule model, which was employed in Ref.\ \cite{Dias:2011mi}
to estimate the mass of the molecule $DD^{\ast }$. Studies there were
performed in the context of the QCD spectral sum rule method by including
into analysis vacuum condensates up to dimension-$6$. The mass of the
molecule $DD^{\ast }$ was found below the two-meson threshold $DD^{\ast } $,
which means that it is bound state and cannot fall apart to a pair of
conventional mesons $D+D^{\ast }.$ Due to closeness of the $DD^{\ast }$
molecule's mass to LHCb data and the same quark content, it may be
interpreted as doubly charmed state $T_{cc}^{+}$. Problem is that, in Ref.\
\cite{Dias:2011mi} width of the molecule $DD^{\ast }$ was not computed.

The masses of the tetraquark $T_{cc}^{+}$ and molecule $DD^{\ast }$ have
been extracted with theoretical uncertainties, which are unavoidable feature
of all sum rule computations. These uncertainties can be reduced only up to
some limits by including into analysis higher dimensional condensates.
Therefore based only on this information it is impossible to distinguish
molecule and tetraquark states from each another. Only way to make strong
statements about internal organization of the $T_{cc}^{+}$, is to analyze
its decay channels and calculate width of this particle. In the present
work, we have carried out such investigation and computed both the mass and
full width of the $T_{cc}^{+}$. Our results support a hypothesis about
diquark-antidiquark structure of the $T_{cc}^{+}$.

Nevertheless, new investigations are necessary to extract additional
information about the $T_{cc}^{+}$. For example, it will be very interesting
to study strong decays of the molecule $DD^{\ast }$ using the sum rule
method and confront results with the LHCb data and theoretical predictions.
Other parameters of the $T_{cc}^{+}$, like the magnetic dipole moment $\mu $
can also provide useful information on this particle \cite{Azizi:2021aib}.

The molecule picture for the $T_{cc}^{+}$ and its decays were considered in
Refs.\ \cite{Meng:2021jnw,Ling:2021bir,Chen:2021vhg} in the context of
alternative approaches. Predictions for the full with of the $T_{cc}^{+}$
obtained in articles \cite{Meng:2021jnw,Ling:2021bir} are rather small
compared with the LHCb data, whereas in Ref.\ \cite{Chen:2021vhg} a nice
agreement with recent measurements was obtained. Moreover the authors of
this work predicted other doubly charmed resonance with parameters $m=3876~%
\mathrm{MeV}$ and $\Gamma =412~\mathrm{keV}$, which should be object of
investigations. As is seen, even in the limits of the same model results for
parameters of the $T_{cc}^{+}$ differ considerably.

Another problem in analyses of the doubly charmed state $T_{cc}^{+}$ is
connected with essential experimental errors in measurements of $\Gamma $.
More accurate data are required to compare different theoretical models. In
other words, physics of the exotic meson $T_{cc}^{+}$ is far from being
finished and is waiting for future experimental and theoretical studies.

\end{document}